\documentclass[cameraready]{Interspeech}

\title{Beyond Deep Learning: Speech Segmentation and Phone Classification with Neural Assemblies}

\author[affiliation={}, orcid=0009-0002-5763-3311]{Trevor}{Adelson}
\author[affiliation={2}, orcid=0000-0001-8492-1787]{Vidhyasaharan}{Sethu}
\author[affiliation={1}, orcid=0000-0003-3806-1493]{Ting}{Dang}

\address{
    $^1$ The University of Melbourne, Australia \\
    $^2$ The University of New South Wales, Australia
}

\email{xtrevad@proton.me, v.sethu@unsw.edu.au, Ting.Dang@unimelb.edu.au}


\usepackage{comment}
\usepackage{soul, xcolor}

\usepackage{tikz}
\usepackage{amsmath} 
\usepackage{amssymb} 
\usetikzlibrary{arrows.meta, positioning, calc, fit, backgrounds, decorations.pathreplacing}
\begin{document}

\maketitle

\begin{abstract}
Deep learning dominates speech processing but relies on massive datasets, backpropagation, and produces dense representations. Assembly Calculus (AC), which models sparse neuronal assemblies with Hebbian plasticity and winner-take-all competition, offers a biologically grounded alternative. We introduce an AC-based speech processing framework that operates directly on continuous speech by combining three key contributions: (i) neural encoding that converts speech into assembly-compatible spike patterns; (ii) a multi-area architecture organising assemblies across hierarchical timescales and classes; and (iii) cross-area update schemes. Applied to two tasks of boundary detection and segment classification, our framework detects phone (F1=0.69) and word (F1=0.61) boundaries without weight training and achieves 47.5\% and 45.1\% accuracy on phone and command recognition. These results show that AC-based dynamical systems are a promising alternative to deep learning for speech.
\end{abstract}

\section{Introduction}
Deep learning (DL) has significantly transformed speech processing over the last two decades with architectures based on convolutional and recurrent neural networks, and more recently Transformer and Conformer models, now being almost universally adopted for tasks such as automatic speech recognition, speaker verification, and paralinguistic analysis. Large-scale self-supervised approaches such as wav2vec 2.0~\cite{baevskiWav2vec20Framework2020} and HuBERT~\cite{hsuHuBERTSelfSupervisedSpeech2021} demonstrate that representations learned from vast amounts of unlabelled audio can be fine-tuned to achieve near–human-level performance on standard benchmarks. 

However, despite these advances and the models exhibiting high accuracies at a number of benchmarks, when compared to the adaptive, efficient, and structured processing exhibited by the human brain, current deep learning systems reveal several important limitations. Chief amongst them is the extremely resource intensive nature of model training . For instance, a state-of-the-art ASR such as Whisper is trained on 680,000 hours of speech data (approximately 77 years of speech)~\cite{radford_robust_2022}, which is orders of magnitude more data than humans need to learn to speak. This limitation in turn is a consequence of other architectural and computational limitations. For example, DL models are trained using stochastic gradient descent which requires all model parameters to be updated during training.
Additionally, the internal representations within DL models are typically not sparse, nor do they lend themselves in any obvious manner to composability of the concepts that are being represented such as phonetic content, prosody, or speaker identity. Finally, DL models are predominantly feed-forward models (with the exception of some limited recurrent structures such as RNNs, LSTMs, etc.), which in turn makes them highly suited for running on GPUs but less ideal for problems that are inherently recurrent such as engaging in a speech conversation with no information about when the interaction might end. Furthermore, adding new phones, acoustic categories, or speaker identities usually requires updating shared parameters, often risking catastrophic forgetting~\cite{Kemker_McClure_Abitino_Hayes_Kanan_2018}. This tight coupling between representations complicates continual learning and makes it difficult to manipulate or recombine learned phonetic and prosodic primitives without retraining large parts of the system. %

In contrast, the human brain learns continuously from limited exposure, without making a distinction between “training” and “inference,” and under strict metabolic power constraints~\cite{Lake2015-kj}~\cite{Bi1998-tk}. Biological learning is inherently incremental and energy-efficient, whereas deep learning pipelines are largely static and resource-intensive. This gap between current deep learning systems and biological intelligence in terms of efficiency, locality, compositionality, and continual adaptation motivates the exploration of alternative computational frameworks inspired more directly by neural principles such as Assembly Calculus (AC), introduced by Papadimitriou et al \cite{papadimitriouBrainComputationAssemblies2020}. AC provides a computational model of how highly recurrent networks of neurons can represent and manipulate information. In this framework, the basic units are sparse groups of neurons, called assemblies, of which only a few are active at any given time. These assemblies form and change according to simple, local learning rules: when two neurons fire together, the synapse between them is strengthened (Hebbian plasticity), and only the most strongly driven neurons in an area are allowed to fire (winner-take-all inhibition). AC has been applied to language parsing~\cite{mitropolskyBiologicallyPlausibleParser2021,weiBionicNaturalLanguage2024}, classification~\cite{dabagiaAssembliesNeuronsLearn2022}, planning~\cite{damorePlanningBiologicalNeurons2022}, and finite state machine simulation~\cite{dabagiaComputationSequencesAssemblies2024}. Because AC employs Hebbian learning with winner-take-all inhibition within local areas, it makes no distinction between training and inference and also sidesteps the requirement to propagate error signals throughout the model.

However, AC has important limitations that must be addressed before it can be applied to speech processing. First, all existing AC work assumes discrete, linearly separable input symbols~\cite{papadimitriouBrainComputationAssemblies2020}, whereas real speech signals are continuous and highly coarticulated in time. It is therefore unclear how to construct a binary neural encoding that maps continuous speech features into assemblies while preserving the temporal and spectral structure needed for phone classification. Second, standard AC does not specify how information should be organised across different temporal and spectral scales. Speech processing requires representations that simultaneously capture ``when'' (e.g., phone or word boundary detection) and ``what'' (category recognition). It remains unclear whether a single assembly-based representation can support both functions, or how coding schemes should differ across hierarchical linguistic levels. Third, AC relies on local plasticity and does not include an explicit global loss function. As a result, there is no direct mechanism for optimising task-level objectives such as boundary F1 or phone classification accuracy. Existing demonstrations are typically limited to small, idealised symbolic sequences with pre-defined assemblies, and do not address how to organise and stabilise many competing assemblies corresponding to large phone or word inventories, nor do they define how class decisions should be read out from evolving assembly trajectories.

In this work, we address these limitations which in turn lets us instantiate AC in concrete, task-driven speech processing systems, targeting two core tasks: \textit{temporal segmentation}, i.e., identifying the onset times of phones and words in continuous speech without supervised boundary labels; and \textit{segment classification}, i.e., assigning a phonetic or word-level category label to a given speech segment. We first introduce a binary neural encoding based on probabilistic mel-spectrogram binarisation and population-coded MFCC representations, enabling continuous speech features to be mapped into assembly-compatible inputs. We then develop refractory assembly hierarchies specialised for boundary detection, allowing assemblies to encode temporal segmentation. For classification, we introduce the idea of per-class recurrent areas and a trajectory-based resonance scoring mechanism to read out phone and word identities from assembly dynamics.
Together, these components provide a principled implementation of AC for continuous speech, opening up a new line of inquiry that moves beyond conventional deep learning approaches towards more biologically inspired models of speech processing.

\section{Assembly Calculus}

\subsection{Framework}

\begin{figure}[t]
\centering
\includegraphics[width=0.9\linewidth]{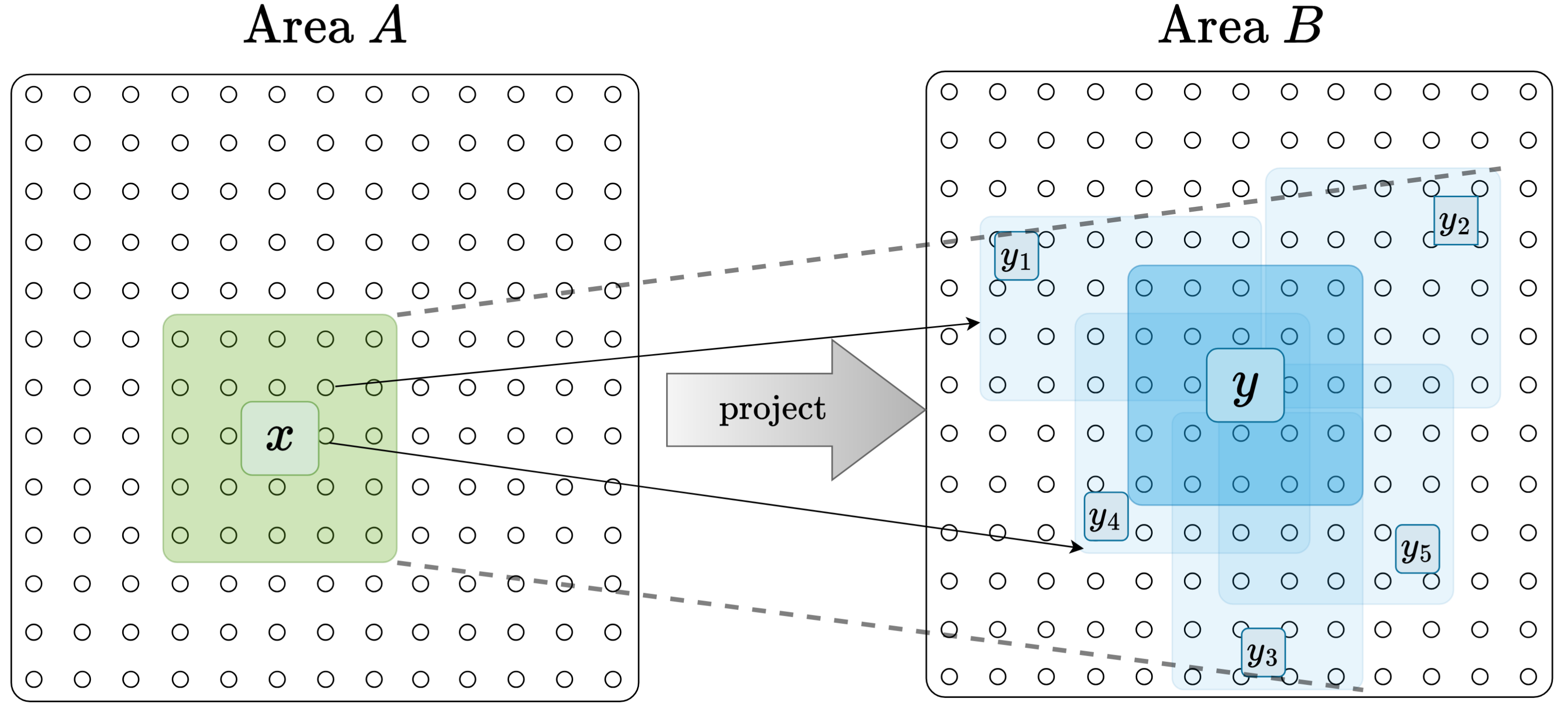}
\caption{Fundamental \textbf{Project} operation of Assembly Calculus. Binary input vectors are projected into a neural area where the $k$-cap operation selects the top-$k$ most activated neurons to form a sparse assembly. Recurrent plasticity strengthens connections between co-active neurons, causing assemblies to stabilise over repeated presentations of the same input.}
\vspace{-10pt}
\label{fig:ac_fundamentals}
\end{figure}

In AC, the `brain' is modelled as a collection of \textit{areas}, each containing $n$ excitatory neurons of which at most $k \ll n$ may fire simultaneously. Areas are connected by sparse random bipartite graphs; we use a fixed in-degree $k_\mathrm{in}$ as a deterministic equivalent of the Erd\H{o}s–R\'{e}nyi model in the sparse limit. At each discrete time step, neuron $v$ computes its total synaptic input $s_v = \sum_{u \in \mathrm{src}(v)} w_{uv}\, x_u$, and $k$ neurons with the largest $s_v$ are selected (the \textit{$k$-cap} operation). The resulting binary activation pattern is the \textit{assembly}: a sparse, distributed code based on the current input (Figure~\ref{fig:ac_fundamentals}).

The key operations are: \textit{projection}, in which stimulating an area through feedforward connections forms a new assembly in a target area; \textit{association}, in which repeated co-activation strengthens shared connections; and \textit{reciprocal stabilisation}, in which bidirectional connections between areas cause assemblies to reinforce each other. Under Erd\H{o}s--R\'{e}nyi random graph assumptions, Papadimitriou et al.\cite{papadimitriouBrainComputationAssemblies2020} showed these operations can: simulate $\mathcal{O}(nk)$ space-bounded computations. Mitropolsky et al.\cite{mitropolskyBiologicallyPlausibleParser2021}
demonstrated that hierarchical syntactic structure can emerge from assembly dynamics alone, implementing a parser for natural language using only AC operations (with pre-defined symbolic assemblies). The parser works by assigning each word a set of gate commands that control which brain areas (SUBJ, VERB, OBJ, etc.) are open for communication. As each word is read in, it projects its assembly into the appropriate open area, and the gating ensures that only grammatically valid connections form, e.g. "man" projects into SUBJ, "saw" projects into VERB while also absorbing the existing SUBJ assembly. The final parse tree is recovered by tracing which assemblies became linked through these projections.

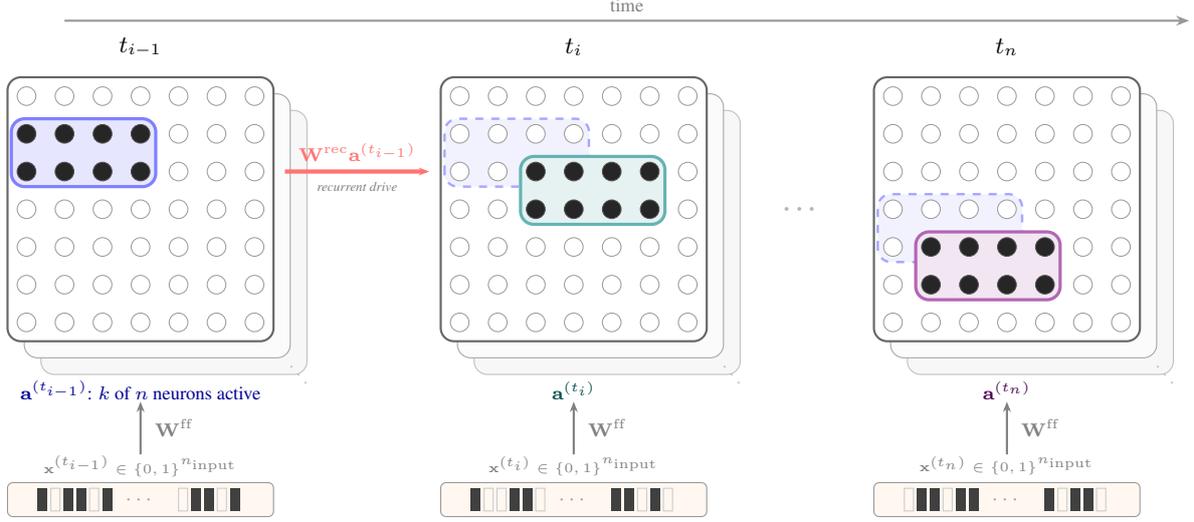
\begin{figure*}[t]
\centering

\tikzset{
  neuron/.style={circle, draw=black!40, minimum size=7pt, inner sep=0pt, line width=0.35pt},
  active/.style={neuron, fill=black!85, draw=black!70},
  inactive/.style={neuron, fill=white},
  areabox/.style={draw=black!60, rounded corners=5pt, thick},
  arr/.style={-{Stealth[length=4pt]}, thick},
  lbl/.style={font=\scriptsize},
  sublbl/.style={font=\tiny\itshape, text=black!50},
}

\begin{tikzpicture}

\draw[-{Stealth[length=5pt]}, thick, black!40] (-1.0, 3.90) -- (13.8, 3.90)
  node[lbl, midway, above, text=black!50] {time};

\node[font=\small\bfseries] at (0, 3.55) {$t_{i{-}1}$};

\draw[rounded corners=5pt, draw=black!25, fill=gray!6]
  (-1.31, -0.79) rectangle (2.19, 2.71);
\node[font=\scriptsize, text=black!25] at (2.08, -0.68) {$\ddots$};
\draw[rounded corners=5pt, draw=black!35, fill=gray!3]
  (-1.53, -0.57) rectangle (1.97, 2.93);
\draw[areabox, fill=white] (-1.75, -0.35) rectangle (1.75, 3.15);

\fill[blue!10, rounded corners=5pt] (-1.70, 1.70) rectangle (0.20, 2.60);
\draw[blue!50, rounded corners=5pt, line width=1.2pt] (-1.70, 1.70) rectangle (0.20, 2.60);

\node[inactive] at (-1.50, 2.90) {};
\node[inactive] at (-1.00, 2.90) {};
\node[inactive] at (-0.50, 2.90) {};
\node[inactive] at ( 0.00, 2.90) {};
\node[inactive] at ( 0.50, 2.90) {};
\node[inactive] at ( 1.00, 2.90) {};
\node[inactive] at ( 1.50, 2.90) {};
\node[active]   at (-1.50, 2.40) {};
\node[active]   at (-1.00, 2.40) {};
\node[active]   at (-0.50, 2.40) {};
\node[active]   at ( 0.00, 2.40) {};
\node[inactive] at ( 0.50, 2.40) {};
\node[inactive] at ( 1.00, 2.40) {};
\node[inactive] at ( 1.50, 2.40) {};
\node[active]   at (-1.50, 1.90) {};
\node[active]   at (-1.00, 1.90) {};
\node[active]   at (-0.50, 1.90) {};
\node[active]   at ( 0.00, 1.90) {};
\node[inactive] at ( 0.50, 1.90) {};
\node[inactive] at ( 1.00, 1.90) {};
\node[inactive] at ( 1.50, 1.90) {};
\node[inactive] at (-1.50, 1.40) {};
\node[inactive] at (-1.00, 1.40) {};
\node[inactive] at (-0.50, 1.40) {};
\node[inactive] at ( 0.00, 1.40) {};
\node[inactive] at ( 0.50, 1.40) {};
\node[inactive] at ( 1.00, 1.40) {};
\node[inactive] at ( 1.50, 1.40) {};
\node[inactive] at (-1.50, 0.90) {};
\node[inactive] at (-1.00, 0.90) {};
\node[inactive] at (-0.50, 0.90) {};
\node[inactive] at ( 0.00, 0.90) {};
\node[inactive] at ( 0.50, 0.90) {};
\node[inactive] at ( 1.00, 0.90) {};
\node[inactive] at ( 1.50, 0.90) {};
\node[inactive] at (-1.50, 0.40) {};
\node[inactive] at (-1.00, 0.40) {};
\node[inactive] at (-0.50, 0.40) {};
\node[inactive] at ( 0.00, 0.40) {};
\node[inactive] at ( 0.50, 0.40) {};
\node[inactive] at ( 1.00, 0.40) {};
\node[inactive] at ( 1.50, 0.40) {};
\node[inactive] at (-1.50, -0.10) {};
\node[inactive] at (-1.00, -0.10) {};
\node[inactive] at (-0.50, -0.10) {};
\node[inactive] at ( 0.00, -0.10) {};
\node[inactive] at ( 0.50, -0.10) {};
\node[inactive] at ( 1.00, -0.10) {};
\node[inactive] at ( 1.50, -0.10) {};

\node[lbl, blue!60!black] at (0, -1.00) {$\mathbf{a}^{(t_{i-1})}$: $k$ of $n$ neurons active};

\draw[arr, black!50] (0, -1.85) -- (0, -1.15)
  node[lbl, right, pos=0.5, xshift=2pt] {$\mathbf{W}^{\mathrm{ff}}$};

\node[sublbl] at (0, -2.00) {$\mathbf{x}^{(t_{i-1})} \in \{0,1\}^{n_\mathrm{input}}$};

\node[draw=black!40, rounded corners=2pt, fill=orange!6,
      minimum width=3.5cm, minimum height=0.45cm] at (0, -2.45) {};
\fill[black!75] (-1.35, -2.60) rectangle ++(0.12, 0.30);
\draw[black!20] (-1.18, -2.60) rectangle ++(0.12, 0.30);
\fill[black!75] (-1.01, -2.60) rectangle ++(0.12, 0.30);
\fill[black!75] (-0.84, -2.60) rectangle ++(0.12, 0.30);
\draw[black!20] (-0.67, -2.60) rectangle ++(0.12, 0.30);
\fill[black!75] (-0.50, -2.60) rectangle ++(0.12, 0.30);
\node[font=\scriptsize, text=black!40] at (0.0, -2.45) {$\cdots$};
\draw[black!20] (0.50, -2.60) rectangle ++(0.12, 0.30);
\fill[black!75] (0.67, -2.60) rectangle ++(0.12, 0.30);
\fill[black!75] (0.84, -2.60) rectangle ++(0.12, 0.30);
\draw[black!20] (1.01, -2.60) rectangle ++(0.12, 0.30);
\fill[black!75] (1.18, -2.60) rectangle ++(0.12, 0.30);

\draw[arr, red!55, line width=1.6pt]
  (1.90, 1.90) -- (3.80, 1.90)
  node[lbl, above, midway, text=red!55] {$\mathbf{W}^{\mathrm{rec}} \mathbf{a}^{(t_{i-1})}$}
  node[sublbl, below, midway] {recurrent drive};

\node[font=\small\bfseries] at (5.7, 3.55) {$t_i$};

\draw[rounded corners=5pt, draw=black!25, fill=gray!6]
  (4.39, -0.79) rectangle (7.89, 2.71);
\node[font=\scriptsize, text=black!25] at (7.78, -0.68) {$\ddots$};
\draw[rounded corners=5pt, draw=black!35, fill=gray!3]
  (4.17, -0.57) rectangle (7.67, 2.93);
\draw[areabox, fill=white] (3.95, -0.35) rectangle (7.45, 3.15);

\fill[blue!5, rounded corners=5pt] (4.00, 1.70) rectangle (5.90, 2.60);
\draw[blue!35, rounded corners=5pt, line width=0.9pt, dashed] (4.00, 1.70) rectangle (5.90, 2.60);

\fill[teal!10, rounded corners=5pt] (5.00, 1.20) rectangle (6.90, 2.10);
\draw[teal!60, rounded corners=5pt, line width=1.2pt] (5.00, 1.20) rectangle (6.90, 2.10);

\node[inactive] at (4.20, 2.90) {};
\node[inactive] at (4.70, 2.90) {};
\node[inactive] at (5.20, 2.90) {};
\node[inactive] at (5.70, 2.90) {};
\node[inactive] at (6.20, 2.90) {};
\node[inactive] at (6.70, 2.90) {};
\node[inactive] at (7.20, 2.90) {};
\node[inactive] at (4.20, 2.40) {};
\node[inactive] at (4.70, 2.40) {};
\node[inactive] at (5.20, 2.40) {};
\node[inactive] at (5.70, 2.40) {};
\node[inactive] at (6.20, 2.40) {};
\node[inactive] at (6.70, 2.40) {};
\node[inactive] at (7.20, 2.40) {};
\node[inactive] at (4.20, 1.90) {};
\node[inactive] at (4.70, 1.90) {};
\node[active]   at (5.20, 1.90) {};
\node[active]   at (5.70, 1.90) {};
\node[active]   at (6.20, 1.90) {};
\node[active]   at (6.70, 1.90) {};
\node[inactive] at (7.20, 1.90) {};
\node[inactive] at (4.20, 1.40) {};
\node[inactive] at (4.70, 1.40) {};
\node[active]   at (5.20, 1.40) {};
\node[active]   at (5.70, 1.40) {};
\node[active]   at (6.20, 1.40) {};
\node[active]   at (6.70, 1.40) {};
\node[inactive] at (7.20, 1.40) {};
\node[inactive] at (4.20, 0.90) {};
\node[inactive] at (4.70, 0.90) {};
\node[inactive] at (5.20, 0.90) {};
\node[inactive] at (5.70, 0.90) {};
\node[inactive] at (6.20, 0.90) {};
\node[inactive] at (6.70, 0.90) {};
\node[inactive] at (7.20, 0.90) {};
\node[inactive] at (4.20, 0.40) {};
\node[inactive] at (4.70, 0.40) {};
\node[inactive] at (5.20, 0.40) {};
\node[inactive] at (5.70, 0.40) {};
\node[inactive] at (6.20, 0.40) {};
\node[inactive] at (6.70, 0.40) {};
\node[inactive] at (7.20, 0.40) {};
\node[inactive] at (4.20, -0.10) {};
\node[inactive] at (4.70, -0.10) {};
\node[inactive] at (5.20, -0.10) {};
\node[inactive] at (5.70, -0.10) {};
\node[inactive] at (6.20, -0.10) {};
\node[inactive] at (6.70, -0.10) {};
\node[inactive] at (7.20, -0.10) {};

\node[lbl, teal!60!black] at (5.7, -1.00) {$\mathbf{a}^{(t_i)}$};

\draw[arr, black!50] (5.7, -1.85) -- (5.7, -1.15)
  node[lbl, right, pos=0.5, xshift=2pt] {$\mathbf{W}^{\mathrm{ff}}$};

\node[sublbl] at (5.7, -2.00) {$\mathbf{x}^{(t_i)} \in \{0,1\}^{n_\mathrm{input}}$};

\node[draw=black!40, rounded corners=2pt, fill=orange!6,
      minimum width=3.5cm, minimum height=0.45cm] at (5.7, -2.45) {};
\fill[black!75] (4.35, -2.60) rectangle ++(0.12, 0.30);
\draw[black!20] (4.52, -2.60) rectangle ++(0.12, 0.30);
\draw[black!20] (4.69, -2.60) rectangle ++(0.12, 0.30);
\fill[black!75] (4.86, -2.60) rectangle ++(0.12, 0.30);
\fill[black!75] (5.03, -2.60) rectangle ++(0.12, 0.30);
\draw[black!20] (5.20, -2.60) rectangle ++(0.12, 0.30);
\node[font=\scriptsize, text=black!40] at (5.7, -2.45) {$\cdots$};
\fill[black!75] (6.20, -2.60) rectangle ++(0.12, 0.30);
\fill[black!75] (6.37, -2.60) rectangle ++(0.12, 0.30);
\draw[black!20] (6.54, -2.60) rectangle ++(0.12, 0.30);
\fill[black!75] (6.71, -2.60) rectangle ++(0.12, 0.30);
\draw[black!20] (6.88, -2.60) rectangle ++(0.12, 0.30);

\node[font=\normalsize, text=black!30] at (8.70, 1.40) {$\boldsymbol{\cdots}$};

\node[font=\small\bfseries] at (11.4, 3.55) {$t_n$};

\draw[rounded corners=5pt, draw=black!25, fill=gray!6]
  (10.09, -0.79) rectangle (13.59, 2.71);
\node[font=\scriptsize, text=black!25] at (13.48, -0.68) {$\ddots$};
\draw[rounded corners=5pt, draw=black!35, fill=gray!3]
  (9.87, -0.57) rectangle (13.37, 2.93);
\draw[areabox, fill=white] (9.65, -0.35) rectangle (13.15, 3.15);

\fill[blue!5, rounded corners=5pt] (9.70, 0.70) rectangle (11.60, 1.60);
\draw[blue!35, rounded corners=5pt, line width=0.9pt, dashed] (9.70, 0.70) rectangle (11.60, 1.60);

\fill[violet!10, rounded corners=5pt] (10.20, 0.20) rectangle (12.10, 1.10);
\draw[violet!60, rounded corners=5pt, line width=1.2pt] (10.20, 0.20) rectangle (12.10, 1.10);

\node[inactive] at (9.90, 2.90) {};
\node[inactive] at (10.40, 2.90) {};
\node[inactive] at (10.90, 2.90) {};
\node[inactive] at (11.40, 2.90) {};
\node[inactive] at (11.90, 2.90) {};
\node[inactive] at (12.40, 2.90) {};
\node[inactive] at (12.90, 2.90) {};
\node[inactive] at (9.90, 2.40) {};
\node[inactive] at (10.40, 2.40) {};
\node[inactive] at (10.90, 2.40) {};
\node[inactive] at (11.40, 2.40) {};
\node[inactive] at (11.90, 2.40) {};
\node[inactive] at (12.40, 2.40) {};
\node[inactive] at (12.90, 2.40) {};
\node[inactive] at (9.90, 1.90) {};
\node[inactive] at (10.40, 1.90) {};
\node[inactive] at (10.90, 1.90) {};
\node[inactive] at (11.40, 1.90) {};
\node[inactive] at (11.90, 1.90) {};
\node[inactive] at (12.40, 1.90) {};
\node[inactive] at (12.90, 1.90) {};
\node[inactive] at (9.90, 1.40) {};
\node[inactive] at (10.40, 1.40) {};
\node[inactive] at (10.90, 1.40) {};
\node[inactive] at (11.40, 1.40) {};
\node[inactive] at (11.90, 1.40) {};
\node[inactive] at (12.40, 1.40) {};
\node[inactive] at (12.90, 1.40) {};
\node[inactive] at (9.90, 0.90) {};
\node[active]   at (10.40, 0.90) {};
\node[active]   at (10.90, 0.90) {};
\node[active]   at (11.40, 0.90) {};
\node[active]   at (11.90, 0.90) {};
\node[inactive] at (12.40, 0.90) {};
\node[inactive] at (12.90, 0.90) {};
\node[inactive] at (9.90, 0.40) {};
\node[active]   at (10.40, 0.40) {};
\node[active]   at (10.90, 0.40) {};
\node[active]   at (11.40, 0.40) {};
\node[active]   at (11.90, 0.40) {};
\node[inactive] at (12.40, 0.40) {};
\node[inactive] at (12.90, 0.40) {};
\node[inactive] at (9.90, -0.10) {};
\node[inactive] at (10.40, -0.10) {};
\node[inactive] at (10.90, -0.10) {};
\node[inactive] at (11.40, -0.10) {};
\node[inactive] at (11.90, -0.10) {};
\node[inactive] at (12.40, -0.10) {};
\node[inactive] at (12.90, -0.10) {};

\node[lbl, violet!60!black] at (11.4, -1.00) {$\mathbf{a}^{(t_n)}$};

\draw[arr, black!50] (11.4, -1.85) -- (11.4, -1.15)
  node[lbl, right, pos=0.5, xshift=2pt] {$\mathbf{W}^{\mathrm{ff}}$};

\node[sublbl] at (11.4, -2.00) {$\mathbf{x}^{(t_n)} \in \{0,1\}^{n_\mathrm{input}}$};

\node[draw=black!40, rounded corners=2pt, fill=orange!6,
      minimum width=3.5cm, minimum height=0.45cm] at (11.4, -2.45) {};
\draw[black!20] (10.05, -2.60) rectangle ++(0.12, 0.30);
\fill[black!75] (10.22, -2.60) rectangle ++(0.12, 0.30);
\fill[black!75] (10.39, -2.60) rectangle ++(0.12, 0.30);
\draw[black!20] (10.56, -2.60) rectangle ++(0.12, 0.30);
\fill[black!75] (10.73, -2.60) rectangle ++(0.12, 0.30);
\fill[black!75] (10.90, -2.60) rectangle ++(0.12, 0.30);
\node[font=\scriptsize, text=black!40] at (11.4, -2.45) {$\cdots$};
\fill[black!75] (11.90, -2.60) rectangle ++(0.12, 0.30);
\draw[black!20] (12.07, -2.60) rectangle ++(0.12, 0.30);
\fill[black!75] (12.24, -2.60) rectangle ++(0.12, 0.30);
\fill[black!75] (12.41, -2.60) rectangle ++(0.12, 0.30);
\draw[black!20] (12.58, -2.60) rectangle ++(0.12, 0.30);

\end{tikzpicture}

\caption{Inside a per-class RecurrentArea $c$ over a period of $n$ frames. $C$ independent areas (one per class) process the same input in parallel. Each neuron receives feedforward drive from the current input $\mathbf{x}^{(t)}$ plus recurrent drive from the previous assembly $\mathbf{a}_{c}^{(t{-}1)}$; the top-$k$ neurons fire ($k$-cap). Plasticity strengthens co-active feedforward and recurrent edges, causing each area to learn class-specific spectro-temporal trajectories. The area with learned dynamics that best match the input (highest resonance score $R_c$) determines the predicted class.}
\vspace{-10pt}
\label{fig:assembly_detail}
\end{figure*}

\subsection{Temporal Dynamics}

Extending AC to sequential processing, Dabagia et al.\cite{dabagiaComputationSequencesAssemblies2024}
 showed that recurrent connections within an area naturally give rise to \textit{synfire chains}: when a sequence of stimuli is presented repeatedly, assemblies form at each position and forward recurrent weights strengthen, resulting in \textit{pattern completion}, i.e., sequence recall from partial input.

A serious challenge for sequential computation is \textit{transition disambiguation}: when the same stimulus recurs at different positions in a sequence, the system must produce different successors depending on context. \textit{Refractory adaptation} \cite{dabagiaComputationSequencesAssemblies2024}
solves this: neurons that fire accumulate an input-proportional negative bias that suppresses their future activation. Sequence learning with refractory adaptation create a discrete-time dynamical system, with the trajectory through assembly space depending on the full history of prior activations, not just the current input.

\subsection{Plasticity Rules}

Weights are updated after each $k$-cap step by a local plasticity rule. In this paper we use standard Hebbian learning~\cite{hebbOrganizationBehaviorNeuropsychological1949}: for each edge $e$ with weight $w_e$ where both pre- and post-synaptic neurons are active,
\begin{equation}\label{eq:hebbian}
    w_e \leftarrow w_e(1{+}\beta)
\end{equation}
where $\beta$ is the learning rate, followed by per-destination normalisation. All weight updates are purely local, depending only on pre- and post-synaptic activity per-edge.

Standard Hebbian learning only strengthens co-active connections, relying entirely on renormalisation to implicitly weaken others. This can cause assemblies to merge over time, as inactive-presynaptic connections to winning neurons are never explicitly penalised. To address this, we additionally employ a variant of the ABS (Artola--Br\"ocher--Singer) plasticity rule ~\cite{Artola1990-gk}~\cite{garagnaniRecruitmentConsolidationCell2009}, which adds \textit{heterosynaptic long-term depression} (LTD): when a postsynaptic neuron fires but a presynaptic neuron did not contribute, that connection is weakened:
\begin{equation}\label{eq:abs}
    w_e \leftarrow \begin{cases}
        w_e(1{+}\beta) & \text{if pre and post active (LTP)} \\
        w_e(1{-}\beta) & \text{if pre inactive, post active (LTD)} \\
        w_e & \text{otherwise}
    \end{cases}
\end{equation}
followed by clamping $w_e \geq 0$ and per-destination normalisation. This competitive mechanism ensures that winning neurons strengthen only the connections that actually drove them, while explicitly weakening irrelevant inputs.

Within AC models, information is encoded in sparse neuronal assemblies that can be created, re-used, and combined into sequences~\cite{dabagiaComputationSequencesAssemblies2024}. This provides a natural route to compositionality: assemblies can stand in for phone- or word-like units, and their flexible recombination can in principle support an open vocabulary. Finally, the core operations of AC, such as projection between areas, association through co-activation, and stabilisation via recurrent feedback, give rise to rich temporal dynamics and sequence completion~\cite{dabagiaComputationSequencesAssemblies2024}. These properties make AC-based models an interesting framework for constructing speech processing systems.

\section{Proposed Methods}
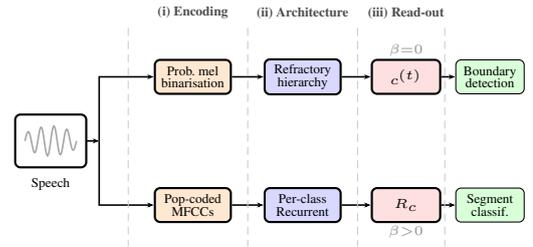
\begin{figure}[t]
\centering
\tikzset{
  encode/.style={draw, rounded corners=2pt, minimum height=0.45cm, minimum width=1.0cm,
               font=\tiny, inner sep=2pt, align=center, fill=orange!18, thick},
  area/.style={draw, rounded corners=2pt, minimum height=0.45cm, minimum width=1.0cm,
               font=\tiny, inner sep=2pt, align=center, fill=blue!15, thick},
  readout/.style={draw, rounded corners=2pt, minimum height=0.45cm, minimum width=0.9cm,
               font=\tiny, inner sep=2pt, align=center, fill=red!12, thick},
  task/.style={draw, rounded corners=2pt, minimum height=0.45cm, minimum width=0.9cm,
               font=\tiny, inner sep=2pt, align=center, fill=green!15},
  arr/.style={-{Stealth[length=2.5pt]}, semithick},
  clabel/.style={font=\tiny\bfseries, text=black!70},
}

\begin{tikzpicture}[x=0.85cm, y=1cm]

\begin{scope}[shift={(0,0)}]
  \draw[thick, rounded corners=2pt] (-0.55,-0.35) rectangle (0.55,0.35);
  \draw[gray!70, semithick] plot[domain=-0.4:0.4, samples=40, smooth]
    (\x, {0.2*sin(1800*\x)*exp(-4*\x*\x)});
  \node[font=\tiny, below] at (0, -0.38) {Speech};
\end{scope}

\coordinate (fork) at (0.75, 0);
\draw[arr] (0.55, 0) -- (fork);

\node[encode] (mel) at (2.2, 0.85) {Prob.\ mel\\[-1pt]binarisation};
\node[area] (refr) at (3.9, 0.85) {Refractory\\[-1pt]hierarchy};
\node[readout] (change) at (5.5, 0.85) {$c^{(t)}$};
\node[task] (bd) at (6.8, 0.85) {Boundary\\[-1pt]detection};

\draw[arr] (fork) |- (mel);
\draw[arr] (mel) -- (refr);
\draw[arr] (refr) -- (change);
\draw[arr] (change) -- (bd);

\node[encode] (mfcc) at (2.2, -0.85) {Pop-coded\\[-1pt]MFCCs};
\node[area] (perclass) at (3.9, -0.85) {Per-class\\[-1pt]Recurrent};
\node[readout] (res) at (5.5, -0.85) {$R_c$};
\node[task] (cl) at (6.8, -0.85) {Segment\\[-1pt]classif.};

\draw[arr] (fork) |- (mfcc);
\draw[arr] (mfcc) -- (perclass);
\draw[arr] (perclass) -- (res);
\draw[arr] (res) -- (cl);

\node[clabel] at (2.2, 1.7) {(i) Encoding};
\node[clabel] at (3.9, 1.7) {(ii) Architecture};
\node[clabel] at (5.5, 1.7) {(iii) Read-out};

\draw[densely dashed, gray!40] (1.2, 1.5) -- (1.2, -1.4);
\draw[densely dashed, gray!40] (3.05, 1.5) -- (3.05, -1.4);
\draw[densely dashed, gray!40] (4.75, 1.5) -- (4.75, -1.4);
\draw[densely dashed, gray!40] (6.1, 1.5) -- (6.1, -1.4);

\node[font=\tiny\itshape, text=black!45] at (5.5, 1.2) {$\beta{=}0$};
\node[font=\tiny\itshape, text=black!45] at (5.5, -1.2) {$\beta{>}0$};

\end{tikzpicture}

\caption{Conceptual overview. Speech is processed by two separate AC pipelines. \textbf{Top:} binarised mel frames drive a frozen-weight refractory hierarchy; the change signal $c^{(t)}$ marks boundaries ($\beta{=}0$). \textbf{Bottom:} population-coded MFCCs drive per-class RecurrentAreas; resonance scoring $R_c$ identifies the class ($\beta{>}0$). Columns mark the three contributions: (i)~neural encoding, (ii)~area architecture, (iii)~task read-out.}
\vspace{-10pt}
\label{fig:conceptual_overview}
\end{figure}

In this section, we introduce the use of AC as a concrete speech processing system. We first define a binary neural interface that transforms continuous acoustic features into sparse, spike-like representations compatible with AC areas, thereby bridging the gap between real-valued speech signals and the discrete assembly dynamics assumed by AC. We then specify the architectural organisation and connectivity of these areas to support two core functions: temporal segmentation and segment classification. For temporal segmentation, we configure a fixed-dynamics hierarchy of refractory assembly areas that expose phone- and word-level boundaries in continuous speech. For segment classification, we introduce plastic, per-class recurrent areas whose learned activation trajectories encode phone and word identities. Taken together, these components form a fully specified AC-based speech processing architecture that can be instantiated on real speech corpora and evaluated on standard boundary detection and classification tasks.
Figure~\ref{fig:conceptual_overview} provides a conceptual overview of the framework, highlighting the three novel contributions and where they sit in the processing pipeline.

\subsection{Binary Neural Encoding for Speech}

AC areas operate on sparse binary vectors~\cite{papadimitriouBrainComputationAssemblies2020}. We therefore require an encoding that converts continuous speech features to ``binary spike trains'' while preserving relevant acoustic structure. To this end, we employ two complementary encodings that emphasise different properties of the signal.

\subsubsection{Probabilistic Mel Encoding}
\label{sec:prob_bin}

\begin{figure}[t]
    \centering
    \includegraphics[width=\columnwidth]{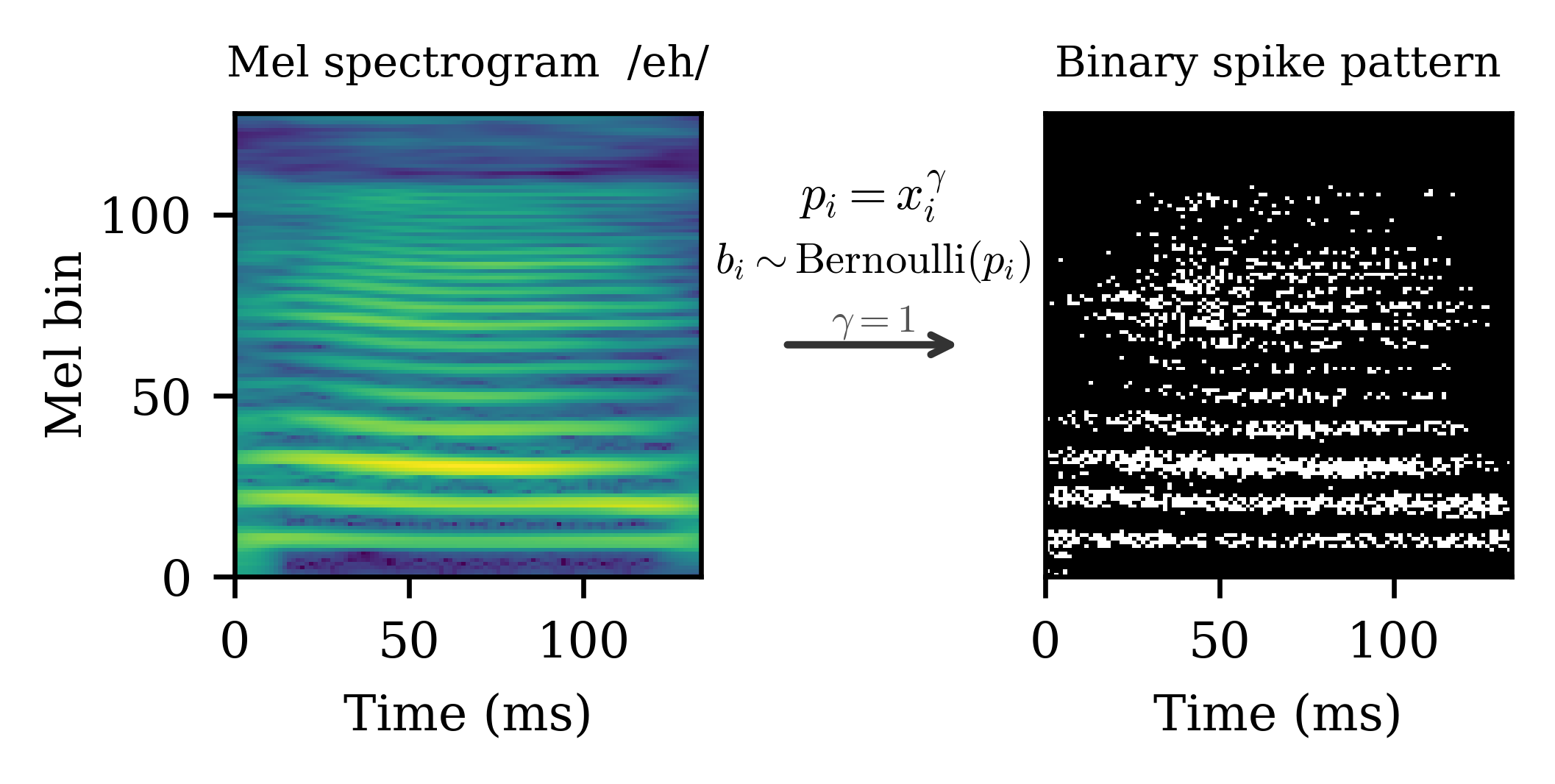}
    \caption{Probabilistic mel binarisation applied to a single phone segment (/eh/, 133\,ms). Left: continuous mel spectrogram normalised to $[0,1]$. Right: binary spike pattern obtained by treating each mel bin value as a Bernoulli firing probability. Brighter spectral regions produce denser activations.}
    \vspace{-10pt}
    \label{fig:prob_bin}
\end{figure}

To preserve fine-grained temporal variation for tasks that require precise boundary recognition, mel-spectral frames are converted into binary vectors through probabilistic sampling, as shown in Figure~\ref{fig:prob_bin}. Given a mel feature vector $\mathbf{x}$, each component $x_i$ is first optionally compressed using a power-law transformation
\begin{equation}
p_i = x_i^{\gamma}, \gamma \le 1
\end{equation}
which reduces dynamic range when $\gamma \le 1$. Each $p_i$ is then interpreted as the firing probability of an input neuron, and a Bernoulli sample is drawn independently for each dimension. As a result, higher-energy mel bins are more likely to activate input neurons, producing sparse stochastic spike patterns that reflect local spectral energy. This encoding preserves frame-to-frame variability in the mel spectrogram and supports dynamic state transitions in downstream AC areas (Figure~\ref{fig:prob_bin}).

\subsubsection{Population-Coded MFCC Encoding}
\label{sec:pop_mfcc}

While probabilistic mel encoding emphasises frame-to-frame change, recognition of quasi-stable acoustic categories (e.g., phones or words) benefits from features that are more invariant to pitch and speaker differences. For this purpose, we use Mel-Frequency Cepstral Coefficients (MFCCs) that are then mapped into sparse binary vectors via Gaussian population coding~\cite{quianquirogaPrinciplesNeuralCoding2013}.

\begin{figure*}[t]
    \centering
    \includegraphics[width=\textwidth]{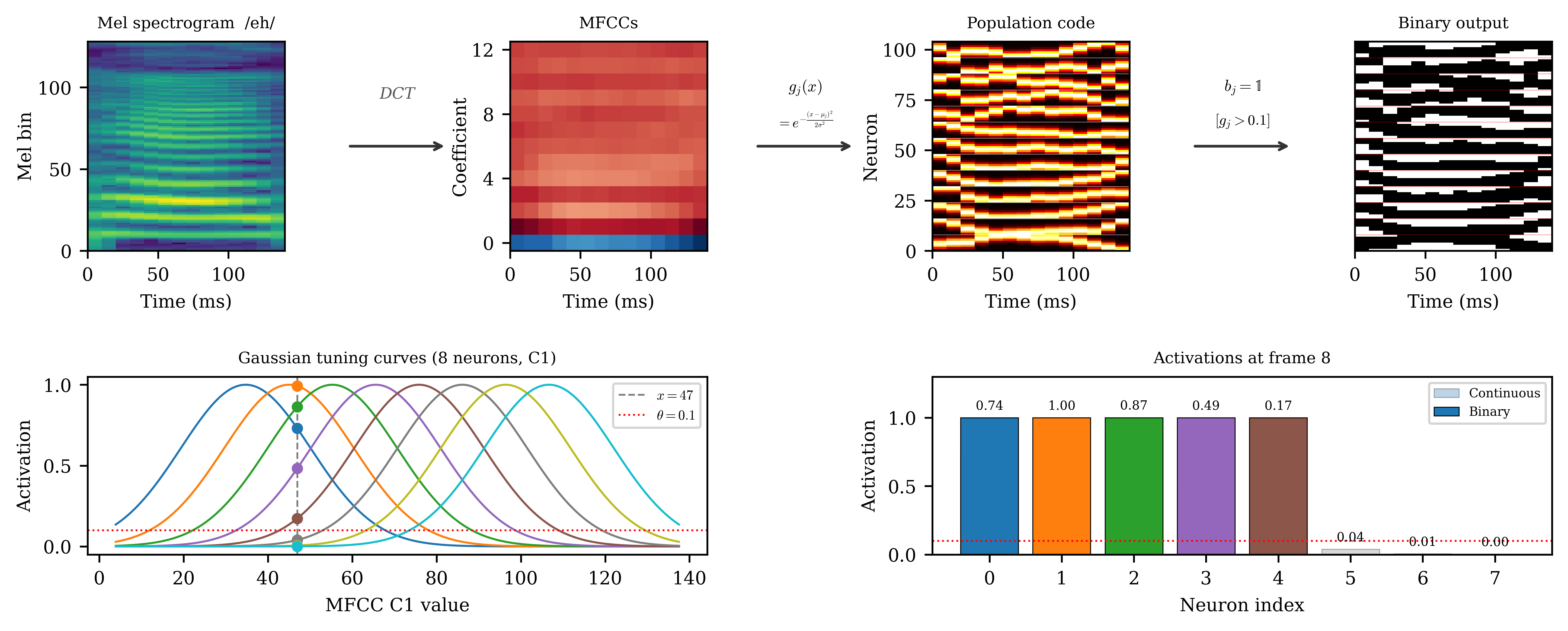}
    \caption{Population-coded MFCC binarisation pipeline. Top row: mel spectrogram of a single phone (/eh/) is transformed into 13 MFCC coefficients, encoded by Gaussian tuning curves into a continuous population code, and thresholded to produce a sparse binary vector. Bottom row: detail of the Gaussian tuning curves for one coefficient (C1), showing how a single MFCC value activates overlapping neurons, and the resulting continuous and binary activations.}
    \vspace{-10pt}
    \label{fig:mfcc_pipeline}
\end{figure*}

As shown in Figure~\ref{fig:mfcc_pipeline}, each of the $M$ MFCC coefficients is represented by a population of $N_{\text{pop}}$ input neurons with Gaussian tuning curves. For coefficient $m$, we first compute the empirical range of its values over the training set and place the preferred values ${\mu_j}_{j=1}^{N{\text{pop}}}$ uniformly between the 1st and 99th percentiles, ensuring that the population tiles the range while being robust to outliers. Given a coefficient value $x$, the activation of neuron $j$ with preferred value $\mu_j$ and width $\sigma$ is: 
\begin{equation}
g_j(x) = \exp\left( -\frac{(x - \mu_j)^2}{2\sigma^2} \right)
\end{equation}

Stacking all $M$ coefficients and their $N_{\text{pop}}$-dimensional responses yields an $M \times N_{\text{pop}}$ continuous population code. We then binarise this code by thresholding each $g_j(x)$, producing a sparse binary input vector of dimensionality $n_{\text{input}} = M \times N_{\text{pop}}$. Neurons whose tuning curves are well-aligned with the current MFCC values become active, while others remain silent. This representation emphasises the overall spectral envelope shape rather than raw energy and yields structured, distributed binary patterns that are well-suited as inputs to AC areas in the classification pathway.

Together, these two encodings provide a biologically plausible binary interface between continuous speech signals and discrete assembly dynamics, supporting sensitivity to fine-grained temporal changes and stable local state representations suitable for higher-level pattern formation respectively.

\subsection{Assembly for Segmentation and Classification}
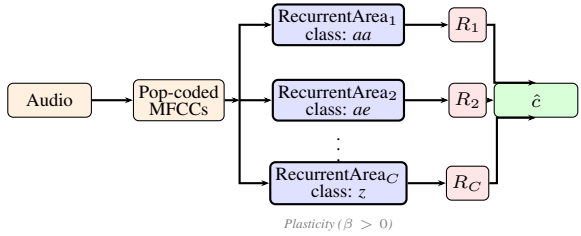
\begin{figure}[t]
\centering

\tikzset{
  area/.style={draw, rounded corners=2pt, minimum height=0.50cm, minimum width=1.3cm,
               font=\scriptsize, inner sep=2pt, align=center, fill=blue!12, thick},
  output/.style={draw, rounded corners=2pt, minimum height=0.45cm, minimum width=1.1cm,
               font=\scriptsize, inner sep=2pt, align=center, fill=green!15},
  proc/.style={draw, rounded corners=2pt, minimum height=0.45cm, minimum width=0.5cm,
               font=\scriptsize, inner sep=2pt, align=center},
  input/.style={draw, rounded corners=2pt, minimum height=0.45cm, minimum width=1.1cm,
               font=\scriptsize, inner sep=2pt, align=center, fill=orange!12},
  score/.style={draw, rounded corners=2pt, minimum height=0.45cm, minimum width=0.45cm,
               font=\scriptsize, inner sep=2pt, align=center, fill=red!10},
  arr/.style={-{Stealth[length=3pt]}, thick},
}

\begin{tikzpicture}[x=1cm, y=1cm]

\node[font=\footnotesize\bfseries, anchor=west] at (-0.2, 1.15) {(a) Hierarchical Boundary Detection};

\node[input] (audio) at (0.4, 0.0) {Audio};

\node[input] (bin) at (2.1, 0.0) {Prob.\ mel\\[-1pt]binarisation};

\node[area] (l1) at (4.0, 0.0) {Level\,1\\[-1pt]\scriptsize Refract.\ Area};

\node[proc] (cs1) at (5.5, 0.0) {$c^{(t)}$};

\node[output] (phn) at (7.0, 0.0) {Phoneme\\[-1pt]boundaries};

\node[area] (l2) at (4.0, -1.0) {Level\,2\\[-1pt]\scriptsize Refract.\ Area};

\node[proc] (cs2) at (5.5, -1.0) {$c^{(t)}$};

\node[output] (wrd) at (7.0, -1.0) {Word\\[-1pt]boundaries};

\draw[arr] (audio) -- (bin);
\draw[arr] (bin) -- (l1);
\draw[arr] (l1) -- (cs1);
\draw[arr] (cs1) -- (phn);
\draw[arr] (l1.south) -- (l2.north) node[font=\tiny, right, xshift=1pt, midway] {$\mathbf{a}^{(t)}$};
\draw[arr] (l2) -- (cs2);
\draw[arr] (cs2) -- (wrd);

\node[font=\tiny\itshape, text=black!50] at (4.0, -1.55) {no learned weights ($\beta=0$)};

\end{tikzpicture}

\vspace{0.3cm}

\begin{tikzpicture}[x=1cm, y=1cm]

\node[font=\footnotesize\bfseries, anchor=west] at (-0.2, 1.95) {(b) Per-Class Phoneme Classification};

\node[input] (audio) at (0.4, 0.0) {Audio};

\node[input] (mfcc) at (2.1, 0.0) {Pop-coded\\[-1pt]MFCCs};

\coordinate (fanout) at (2.9, 0.0);

\node[area] (a1) at (4.2, 1.0) {RecurrentArea$_1$\\[-1pt]\scriptsize class: \textit{aa}};
\node[area] (a2) at (4.2, 0.0) {RecurrentArea$_2$\\[-1pt]\scriptsize class: \textit{ae}};
\node[font=\scriptsize] at (4.2, -0.55) {$\vdots$};
\node[area] (aC) at (4.2, -1.1) {RecurrentArea$_C$\\[-1pt]\scriptsize class: \textit{z}};

\node[score] (r1) at (5.9, 1.0) {$R_1$};
\node[score] (r2) at (5.9, 0.0) {$R_2$};
\node[score] (rC) at (5.9, -1.1) {$R_C$};

\node[output] (argmax) at (6.8, 0.0) {$\hat{c}$};

\draw[arr] (audio) -- (mfcc);
\draw[arr] (mfcc) -- (fanout);

\draw[arr] (fanout) |- (a1.west);
\draw[arr] (fanout) -- (a2.west);
\draw[arr] (fanout) |- (aC.west);

\draw[arr] (a1) -- (r1);
\draw[arr] (a2) -- (r2);
\draw[arr] (aC) -- (rC);

\draw[arr] (r1.east) -- ++(0.1,0) |- (argmax.north);
\draw[arr] (r2) -- (argmax);
\draw[arr] (rC.east) -- ++(0.1,0) |- (argmax.south);

\node[font=\tiny\itshape, text=black!50] at (4.2, -1.65) {Plasticity ($\beta>0$)};

\end{tikzpicture}

\caption{Architecture overview. \textbf{(a)}~Boundary detection: probabilistically binarised mel frames are processed by two cascaded refractory areas; the assembly change signal $c^{(t)}$ marks phoneme and word boundaries without any learned weights. \textbf{(b)}~Classification: population-coded MFCC frames are fed in parallel to $C$ independent per-class RecurrentAreas trained with Hebbian plasticity; the resonance score $R_c$ (Eq.~\ref{eq:resonance}) determines the predicted class $\hat{c} = \arg\max_c R_c$.}
\vspace{-15pt}
\label{fig:architecture}
\end{figure}

Figure~\ref{fig:architecture} depicts how the two binary feature streams described in the previous section drive the dynamics of AC areas configured for our two target tasks: temporal segmentation and segment classification.
We employ two specialised area configurations. A \textit{refractory area} is an assembly area augmented with refractory suppression (Section~2.2) and operated without plasticity ($\beta{=}0$); its dynamics are driven entirely by feedforward input and refractory adaptation, making it sensitive to input changes. A \textit{per-class recurrent area} is a RecurrentArea dedicated to a single class and trained with plasticity ($\beta{>}0$); its learned recurrent weights encode class-specific temporal trajectories.
For temporal segmentation, binarised mel frames are processed by a hierarchy of refractory areas whose state changes expose phone- and word-level boundaries in continuous speech. For segment classification, population-coded MFCC sequences are presented to per-class recurrent areas whose learned assembly trajectories act as dynamical templates for phone and word categories. These constructions specify how binary inputs from the neural encoding are transformed into task-relevant assembly dynamics in our AC-based speech system.

\subsubsection{Temporal Segmentation from Refractory Assembly Areas}
Temporal segmentation is formulated as an unsupervised boundary detection task, where the objective is to identify phone and word onset times in continuous speech.

We construct a hierarchy of two refractory areas that operate at different temporal scales and require no learned weights.
For a single area that consists of a population of neurons with recurrent connectivity and a $k$-cap operation, we obtain a sparse binary activity vector $\mathbf{a}^{(t)}$ with exactly $k$ active units. In addition, each neuron is subject to refractory suppression, meaning that recently active neurons are temporarily inhibited. This prevents the same set of neurons from persisting indefinitely and makes the area sensitive to changes in the input.

When synaptic plasticity is disabled ($\beta = 0$), the area functions purely as a dynamical system driven by three mechanisms: feedforward input, recurrent excitation among currently active neurons, and refractory adaptation. Given a sequence of sparse input frames $\{\mathbf{x}^{(t)}\}$, the area produces a trajectory of assemblies $\{\mathbf{a}^{(t)}\}$. If the input remains stable, the winning set of neurons tends to overlap strongly across consecutive frames. When the input distribution changes sufficiently, the competitive dynamics select a different set of $k$ neurons, resulting in a rapid reconfiguration of the assembly.

\noindent\textbf{Phone-scale segmentation (Level 1).}
The Level~1 area receives frame-level binary inputs obtained from probabilistic mel binarisation. Through $k$-cap competition, even moderate shifts in input statistics can alter which neurons win the competition. Refractory suppression further amplifies this sensitivity by discouraging persistence of the previous assembly. As a result, phone transitions produce noticeable reconfigurations in $\mathbf{a}^{(t)}$.

\noindent\textbf{Word-scale segmentation (Level 2).}
The Level~2 area operates on the activity of Level~1 rather than directly on acoustic input. It receives $\mathbf{a}^{(t)}$ as its input stream and applies the same competitive and refractory dynamics, but effectively integrates information over a longer timescale. Because words consist of structured sequences of phone-level assemblies, Level~2 becomes sensitive to larger-scale shifts in the pattern of Level~1 activity. When a word boundary occurs, the distribution of Level~1 assemblies changes more substantially, leading to a stronger reconfiguration at Level~2.

\noindent\textbf{Boundary signal.}
To quantify assembly reconfiguration at either level, we define the change measure:
\begin{equation}
c^{(t)} = 1 - \frac{\mathbf{a}^{(t)} \cdot \mathbf{a}^{(t-1)}}{k}
\label{eq:change}
\end{equation}
Since the dot product counts the number of shared active neurons between consecutive time steps, $c^{(t)}$ measures the fraction of neurons that have changed. Values near zero indicate stable assemblies, while large values indicate abrupt reorganisation. Peaks in $c^{(t)}$ therefore serve as intrinsic boundary indicators.
\begin{figure}[t]
    \centering
    \includegraphics[width=0.9\columnwidth]{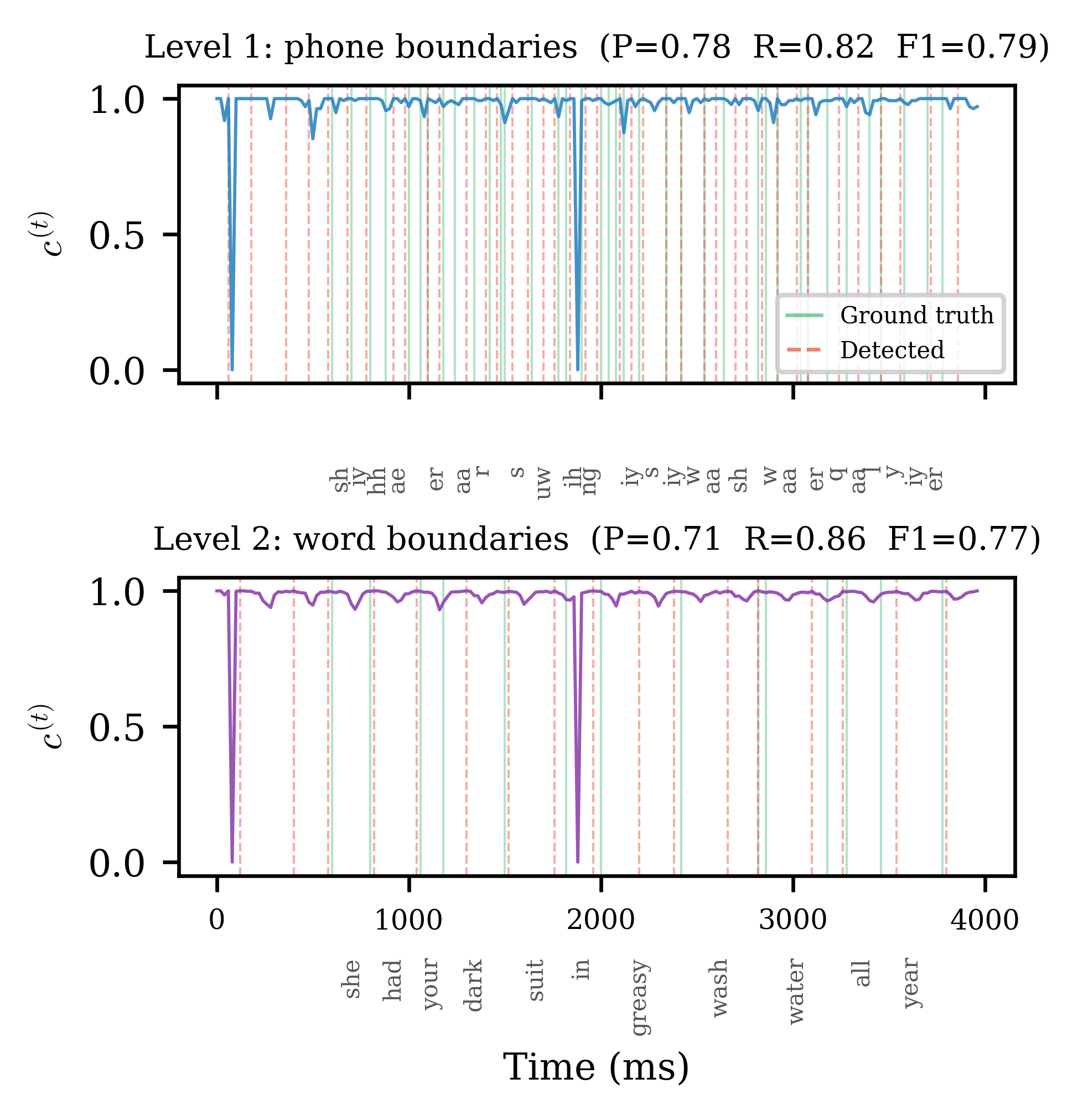}
    \caption{Assembly change signal $c^{(t)}$ (Eq.~\ref{eq:change}) for a single TIMIT utterance. \textbf{Top:} Level~1 (phone boundaries). \textbf{Bottom:} Level~2 (word boundaries). Green vertical lines mark ground-truth boundaries; red dashed lines mark detected peaks. Phone/word labels are shown below each axis.}
    \vspace{-10pt}
    \label{fig:boundary_signal}
\end{figure}

Importantly, both phone- and word-level segmentation emerge without supervised boundary labels or gradient-based optimisation. They arise directly from sparse competition and refractory adaptation operating at different hierarchical levels.

\subsubsection{Segment Classification from Trajectory Scoring}
While temporal segmentation operates in a non-plastic regime and detects boundaries through assembly reconfiguration, segment classification uses the same competitive assembly dynamics in a plastic regime to learn class-specific temporal structure.

We formulate phone classification as a supervised segment classification task using a bank of class-specific recurrent assembly areas. Concretely, we instantiate one \emph{RecurrentArea} per phone class, yielding $C$ independent dynamical systems. Each area learns the characteristic spectro-temporal trajectory of a single phone class.
Acoustic input is represented using population-coded mel-frequency cepstrum coefficients (MFCCs)~\cite{Davis80-COP} (Section~\ref{sec:pop_mfcc}), producing a sparse high-dimensional binary vector $\mathbf{x}^{(t)}$ at each frame.

\noindent\textbf{Recurrent assembly dynamics.}
Each RecurrentArea consists of a neuronal population governed by the same $k$-cap competition principle used in boundary detection. At each time step, neurons receive:
(i) feedforward input from the current acoustic frame via weights $\mathbf{W}^{\mathrm{ff}}$, and
(ii) recurrent input from the previous assembly via weights $\mathbf{W}^{\mathrm{rec}}$.

The total pre-competition input to area $c$ at time $t$ is
\begin{equation}
\mathbf{u}_c^{(t)} =
\mathbf{W}_c^{\mathrm{ff}} \mathbf{x}^{(t)}
+
\mathbf{W}_c^{\mathrm{rec}} \mathbf{a}_c^{(t-1)},
\end{equation}
As before, a $k$-cap rule selects the $k$ neurons with largest total input to form the active assembly $\mathbf{a}_c^{(t)}$. Thus, classification relies on the same sparse competitive mechanism as segmentation; the difference lies in the presence of synaptic plasticity.

\noindent\textbf{Plastic learning of phone dynamics.}
When plasticity is enabled, area $c$ is exposed only to contiguous segments belonging to phone class $c$. Segments are presented in temporal order, allowing the area to learn not just static spectral patterns but their characteristic evolution across time.

Between segments, activations are reset so that each segment begins from a neutral state. This mirrors the ``cold start'' assumption in segmentation and prevents cross-segment interference.

Through repeated exposure, feedforward weights strengthen connections from frequently co-occurring input features, while recurrent weights reinforce transitions between successive assemblies. Consequently, each area internalises a dynamical signature of its phone class: feedforward structure encodes typical spectral evidence, and recurrent structure encodes the typical progression of that evidence.

\noindent\textbf{Resonance-based trajectory scoring}
\label{sec:traj_scoring}
At test time, plasticity is disabled, and a candidate segment is sequentially presented to all $C$ areas. Each area generates an internal assembly trajectory driven by its learned weights.

If the input trajectory aligns with the learned class-specific dynamics of area $c$, feedforward and recurrent inputs reinforce one another over time. This produces consistently strong pre-competition activation values. If the trajectory is inconsistent, recurrent support is weaker and activation remains lower.
\begin{figure*}[t]
\centering
\includegraphics[width=0.9\textwidth]{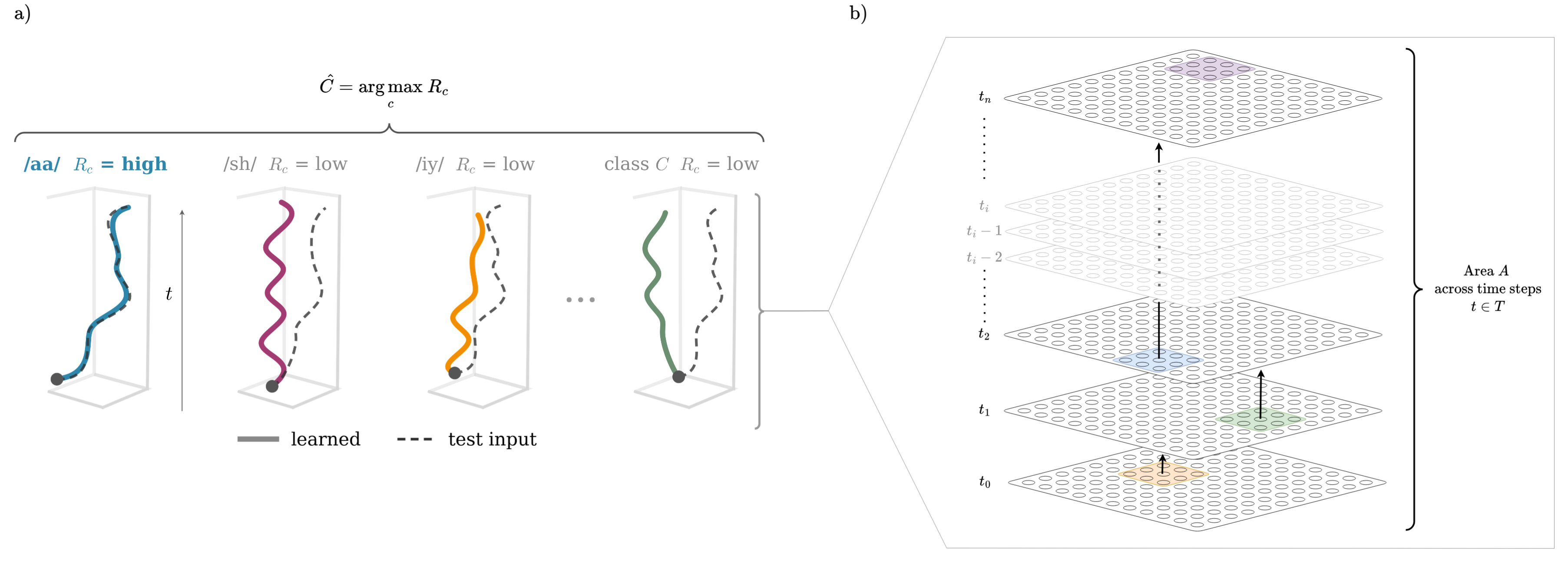}
\caption{Trajectory-based resonance scoring. Each of $C$ per-class RecurrentAreas (a) has learned a characteristic trajectory through assembly space (solid, coloured). The same test input (dashed) is presented to all areas in parallel. The inset (b) shows the frame-by-frame operation: feedforward drive from the current input combines with recurrent drive from the previous assembly, and the top-$k$ neurons fire. The predicted class is the label of the area with the learned trajectory that most closely matches the input, as measured by the resonance score $R_c$.}
\vspace{-15pt}
\label{fig:trajectory}
\end{figure*}

We quantify this alignment using a resonance score:
\begin{equation}\label{eq:resonance}
    R_c = \frac{1}{T'}
    \sum_{t}
    \sum_{i=1}^{k}
    \mathrm{topk}_i\bigl(\mathbf{u}_c^{(t)}\bigr),
\end{equation}
where $T'$ is the number of frames in the segment and $\mathrm{topk}_i(\mathbf{u})$ denotes the $i$-th largest element of $\mathbf{u}$, so the inner sum computes the total pre-competition activation of the $k$ winning neurons. This score measures the cumulative amplification of the input trajectory by the area's learned feedforward and recurrent structure. 
The predicted label is
\begin{equation}
    \hat{c} = \arg\max_c R_c.
\end{equation}

Both segmentation and classification arise from the same underlying assembly dynamics. In segmentation, abrupt changes in assembly configuration (quantified by Eq.~\ref{eq:change}) signal temporal boundaries in a non-plastic regime. In classification, plasticity shapes recurrent structure so that entire trajectories resonate within a class-specific dynamical system. 

Thus, boundaries are detected through transient instability of assemblies, whereas classes are identified through sustained dynamical alignment. In both cases, computation emerges from sparse competition and temporal recurrence rather than gradient-based sequence modeling (Figure~\ref{fig:trajectory}).

\section{Experimental Setup}

\subsection{Dataset}

We evaluate on two standard speech datasets. \textbf{TIMIT}~\cite{TIMIT92} is a corpus of continuous read speech with time-aligned phone and word transcriptions. We use the standard test partition (24~speakers) and 39 phone classes~\cite{Lee89}. %
For boundary detection, we use 20 test utterances. For phone classification, we use 200 training and 50 test utterances. \textbf{Google Speech Commands}~\cite{Warden18} is a dataset of single-word utterances from thousands of speakers. We use the 10-word subset (\textit{yes}, \textit{no}, \textit{up}, \textit{down}, \textit{left}, \textit{right}, \textit{on}, \textit{off}, \textit{stop}, \textit{go}) with 200 training and 50 test samples per class. This evaluates whether per-class trajectory scoring generalises from phone-level segments to whole-word classification with greater speaker and duration variability. We intentionally use a limited dataset to demonstrate the advantages of AC in data-efficient modeling.

\subsection{Feature Extraction}
\textbf{Boundary detection.} Audio at 16\,kHz is represented as a 32-bin log-mel spectrogram (512-sample FFT, 320-sample hop / 20\,ms frames), normalised to $[0,1]$. Frames are binarised via probabilistic sampling: each mel bin value is raised to power $\gamma{=}0.5$ (compressing dynamic range) and rescaled so that 10\% of bins are active across the entire sequence. Each frame is independently Bernoulli-sampled, producing a 32-dimensional binary vector per frame. All binarisation hyperparameters were selected by Bayesian optimisation (Section~\ref{sec:tuning}).

\noindent \textbf{Phone classification.} Audio is represented as $M{=}11$ MFCCs (FFT size 512, hop length 480). Each coefficient is encoded by $N_\mathrm{pop}{=}14$ Gaussian population neurons with ranges set to the 1st--99th percentile of training data, then binarised with threshold~0.044, producing a $154$-dimensional binary vector per frame.

\noindent \textbf{Word classification.} Audio is encoded as $M{=}19$ MFCCs (FFT size 2048, hop length 640). Each coefficient is encoded by $N_\mathrm{pop}{=}9$ Gaussian population neurons and binarised with threshold~0.035, producing a $171$-dimensional binary vector per frame.

\subsection{Architecture Parameters}

\textbf{Level~1 (phone boundary detection):} Frozen-repeat refractory area with $n{=}1{,}531$ neurons, cap $k{=}135$ (8.8\% sparsity), in-degree $k_\mathrm{in}{=}32$, refractory rate $\rho{=}0.989$, similarity threshold $\tau{=}0.761$, minimum peak distance $d{=}3$ frames.

\noindent\textbf{Level~2 (word boundary detection):} Frozen-repeat refractory area with $n{=}6{,}557$ neurons, cap $k{=}588$ (9.0\% sparsity), $k_\mathrm{in}{=}1{,}513$ (near-full connectivity to Level~1), refractory rate $\rho{=}0.092$, similarity threshold $\tau{=}0.745$, minimum peak distance $d{=}7$ frames. 
No Hebbian weight updates are performed during boundary detection inference; all feedforward weights remain at their random initialisation. The change signal (Eq.~\ref{eq:change}) is normalised to $[0,1]$ per utterance.

\noindent\textbf{Per-class RecurrentAreas (phone classification):} 39 independent RecurrentAreas (one per non-silence phone class), each with $n{=}2{,}250$ neurons, cap $k{=}732$ (32.5\% sparsity), in-degree $k_\mathrm{in}{=}87$, recurrent in-degree $k_\mathrm{in}^\mathrm{rec}{=}1{,}245$, learning rate $\beta{=}3.1{\times}10^{-4}$, plasticity rule: ABS (Eq.~\ref{eq:abs}), trained for 13 epochs on contiguous segments.

\noindent\textbf{Per-class RecurrentAreas (word classification):} 10 independent RecurrentAreas (one per word class), each with $n{=}4{,}655$ neurons, cap $k{=}815$ (17.5\% sparsity), in-degree $k_\mathrm{in}{=}74$, recurrent in-degree $k_\mathrm{in}^\mathrm{rec}{=}1{,}237$ (independent mode), learning rate $\beta{=}3.6{\times}10^{-3}$, plasticity rule: ABS, trained for 2 epochs.

\subsection{Hyperparameter Optimisation}
\label{sec:tuning}

All hyperparameters, including feature extraction, binarisation, and area parameters, are jointly optimised using Bayesian optimisation with the Tree-structured Parzen Estimator (TPE) sampler~\cite{Bergstra11} via Optuna~\cite{Akiba19}. For boundary detection, Level~1 parameters (9 hyperparameters spanning mel spectrogram settings, binarisation, and area configuration) are optimised first over 200 trials; Level~2 parameters (6 hyperparameters) are then optimised over 200 trials using the best Level~1 configuration. Classification hyperparameters (${\sim}$20 per task) are optimised analogously over 200 trials.

\subsection{Evaluation}\label{sec:eval}

We evaluate boundary detection using precision, recall, and F1 score. A detected boundary is a true positive if it falls within a tolerance window of a ground-truth boundary: $\pm 2$ frames (40\,ms) for phone boundaries and $\pm 5$ frames (100\,ms) for word boundaries. Boundaries are detected as peaks in the normalised change signal %
with a prominence threshold swept over $\{0.02, 0.05, 0.1, 0.15, 0.2, 0.3, 0.5\}$, selecting the threshold that maximises F1 per utterance. This oracle selection provides an upper bound; a globally fixed threshold reduces F1 by approximately 0.05--0.10. Results are averaged over 10 random seeds to account for the stochastic binarisation.

For phone classification, each per-class RecurrentArea is trained on contiguous segments from 200 TIMIT training utterances (Section~\ref{sec:traj_scoring}). For word classification, each per-class RecurrentArea is trained on 200 whole-word utterances per class from Google Speech Commands. In both cases, segments are classified by highest resonance score $R_c$ (Eq.~\eqref{eq:resonance}) across the $C$ per-class areas.

\section{Results and Discussion}

\subsection{Boundary Detection}

\begin{table}[t]
\caption{Boundary detection results (frozen weights, $\beta{=}0$) on 20 TIMIT test utterances, averaged over 10 random seeds. Prominence threshold is oracle-selected per utterance (Section~\ref{sec:eval}).}
\label{tab:boundary}
\centering
\small
\begin{tabular}{@{}lccc@{}}
\toprule
 & \textbf{Precision} & \textbf{Recall} & \textbf{F1} \\
\midrule
Level~1 (phone) & 0.67 & 0.74 & \textbf{0.69} \\
Level~2 (word) & 0.51 & 0.80 & \textbf{0.61} \\
\midrule
L1-direct (word baseline) & 0.27 & 0.94 & 0.42 \\
\bottomrule
\end{tabular}
\vspace{-10pt}
\end{table}

The Level~1 area achieves a phone boundary F1 score of 0.69, while the hierarchical Level~2 area achieves a word boundary F1 score of 0.61. This improves substantially over using the Level~1 change signal directly at the word level (F1$=$0.42), a gain of $+$0.19 (Table\ref{tab:boundary}). This demonstrates the effectiveness of the proposed framework for boundary detection across different timescales. Optimisation reveals two main trends. First, using a coarser temporal resolution and fewer mel bins reduces variability within phones, which in turn lowers false boundary detections. Second, Level~1 and Level~2 operate on different timescales: Level1 needs a very high refractory rate ($\rho{=}0.989$) to strongly suppress rapid re-firing, whereas Level2 uses a much lower refractory rate ($\rho{=}0.092$) and a larger minimum peak distance ($d{=}7$ vs.\ 3 frames) to reflect the longer duration of words.

\subsection{Classification}

\begin{table}[t]
\caption{Classification results with per-class RecurrentAreas trained using ABS plasticity (Eq.~\ref{eq:abs}) and scored by trajectory resonance (Eq.~\ref{eq:resonance}).}
\vspace{-10pt}
\label{tab:classification}
\centering
\small
\begin{tabular}{@{}lcc@{}}
\toprule
 & \textbf{Phone} & \textbf{Word} \\
 & (TIMIT) & (Speech Cmds) \\
\midrule
Classes $C$ & 39 & 10 \\
Chance accuracy & 2.6\% & 10.0\% \\
Test samples & 4{,}328 & 2{,}000 \\
\midrule
MFCCs $M$ & 11 & 19 \\
Pop.\ neurons $N_\mathrm{pop}$ & 14 & 9 \\
Input dim.\ $n_\mathrm{input}$ & 154 & 171 \\
Area size $n$ & 2{,}250 & 4{,}655 \\
Cap $k$ (\% sparsity) & 732 (32.5\%) & 815 (17.5\%) \\
Learning rate $\beta$ & $3.1{\times}10^{-4}$ & $3.6{\times}10^{-3}$ \\
Epochs & 13 & 2 \\
\midrule
\textbf{Accuracy} & \textbf{47.5\%} & \textbf{45.1\%} \\
\bottomrule
\end{tabular}
\vspace{-10pt}
\end{table}

Using per-class RecurrentAreas trained on population-coded MFCCs (Section~\ref{sec:traj_scoring}), we evaluate classification on both TIMIT phones and Google Speech Commands words (Table~\ref{tab:classification}).

\begin{figure}[t]
    \centering
    \includegraphics[width=0.8\columnwidth]{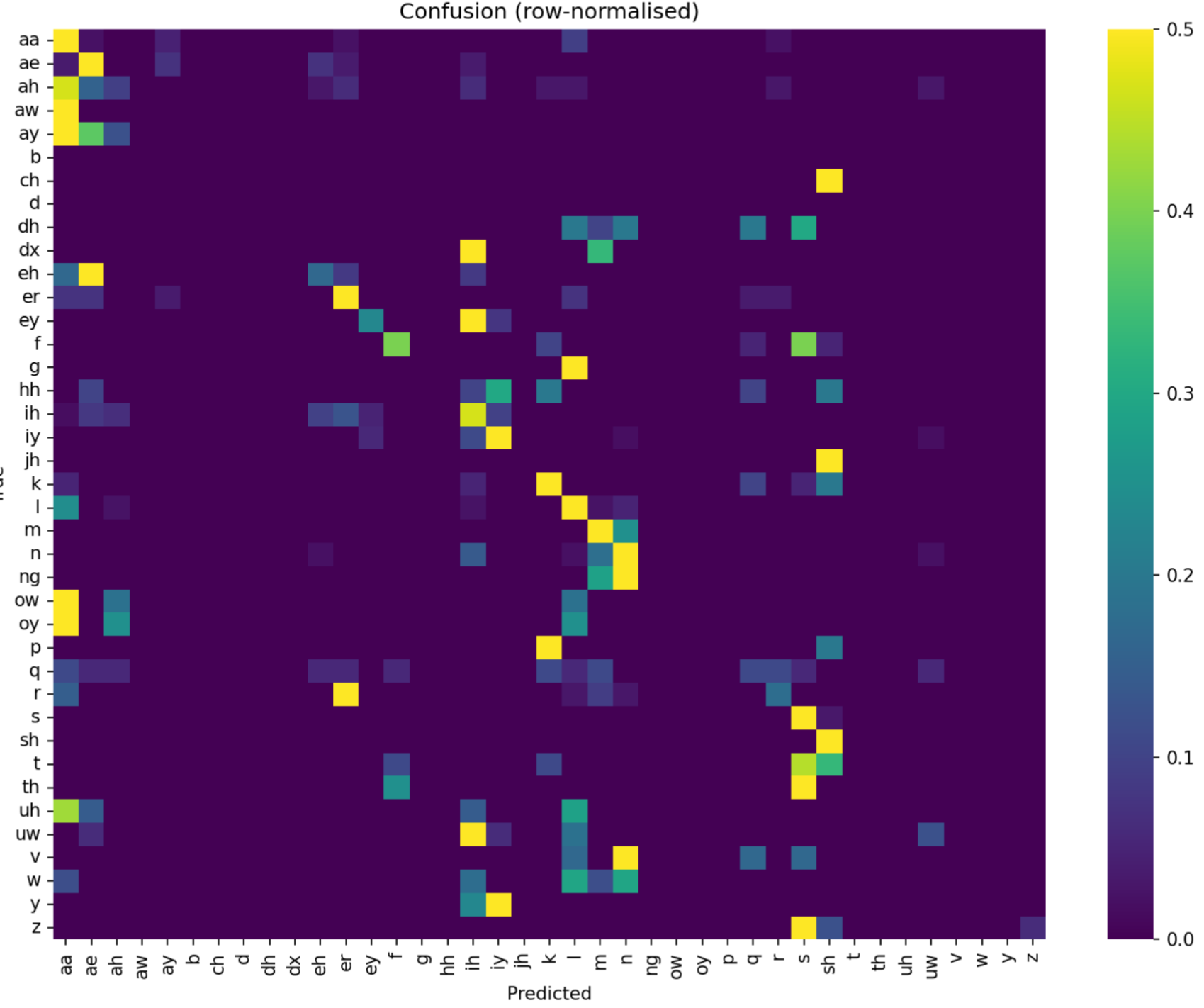}
    \vspace{-10pt}
    \caption{Row-normalised confusion matrix for phone classification on TIMIT (39 classes). Confusions follow phonetic similarity: fricatives (\textit{s}, \textit{sh}, \textit{th}, \textit{z}), low vowels (\textit{aa}, \textit{ah}, \textit{aw}), and nasals (\textit{m}, \textit{n}, \textit{ng}) form visible off-diagonal clusters.}
    \vspace{-10pt}
    \label{fig:phoneme_cm}
\end{figure}

On TIMIT, the system classifies ground-truth-bounded segments across 39 phone classes, achieving \textbf{47.5\%} accuracy (chance 2.6\%). On Google Speech Commands, it classifies whole-word utterances across 10 classes, achieving \textbf{45.1\%} accuracy (chance 10.0\%). Both tasks use ABS plasticity (Eq.~\eqref{eq:abs}) and resonance scoring. As shown in Figure~\ref{fig:phoneme_cm}, confusions predominantly follow phonetic similarity: fricatives cluster (\textit{s}$\leftrightarrow$\textit{sh}$\leftrightarrow$\textit{th}$\leftrightarrow$\textit{z}), low vowels cluster (\textit{aa}$\leftrightarrow$\textit{ah}$\leftrightarrow$\textit{aw}$\leftrightarrow$\textit{ay}), and nasals cluster (\textit{m}$\leftrightarrow$\textit{n}$\leftrightarrow$\textit{ng}). The system discriminates well between broad phonetic categories (vowels vs.\ fricatives vs.\ nasals) but struggles within categories where spectral envelopes overlap. This suggests that the population-coded MFCC representation preserves phonetic similarity structure and the per-class recurrent dynamics capture category-level spectral signatures well but struggle with fine intra-category distinctions. Stops (\textit{b}, \textit{d}, \textit{g}, \textit{k}, \textit{p}) remain the most challenging, possibly because their brief, transient spectra provide insufficient frames for the trajectory scoring to accumulate a discriminative signal.

\subsection{Discussion}

These results demonstrate that refractory adaptation in sparse assembly areas can serve as an unsupervised temporal boundary detector for speech. The key insight is that refractory suppression converts temporal change in the input into spatial change in the assembly: when the spectral content shifts (e.g.\ at a phone boundary), previously-active neurons are suppressed, forcing recruitment of a different assembly. The magnitude of this spatial change, measured by assembly overlap, provides a continuous boundary signal without any training. 

The hierarchical application of this mechanism mirrors the temporal hierarchy of speech: phone boundaries occur every 50--100\,ms, while word boundaries occur every 200--500\,ms. By cascading two refractory areas with different dynamics, the system naturally separates these timescales. 

For classification, per-class assembly areas enable the system to learn class-specific spectro-temporal trajectories. Each area operates as an independent dynamical system with its own attractor landscape shaped by Hebbian plasticity. A correctly matched area produces constructive interference between its learned dynamics and the input trajectory, while mismatched areas produce destructive interference as the input conflicts with the stored attractor structure. This is fundamentally different from deep learning, where classification is a static mapping from a final hidden state; here, the temporal dynamics \textit{are} the representation. The recurrent connections implicitly capture spectral dynamics (analogous to delta features in classical ASR~\cite{Rabiner89-ATO}) without explicitly computing derivatives; the recurrence \textit{is} the temporal derivative.

Interestingly, the two tasks favour different input representations: boundary detection works best with probabilistically binarised mel spectrograms, while classification works best with population-coded MFCCs. This suggests the representations capture complementary aspects of the signal: mel spectrograms preserve the fine temporal change structure needed for boundary detection, while MFCCs provide the pitch-invariant spectral identity needed for class discrimination.

\section{Conclusion}
The proposed AC-based dynamical system demonstrates that sparse assembly dynamics can effectively address two core speech tasks, unsupervised segmentation and supervised phone classification, directly from continuous audio, without backpropagation or the need to learn dense representations. By integrating probabilistic neural encodings, the hierarchical refractory areas, and trajectory-based classification, the framework employs simple local plasticity rules to learn stable assemblies that both detect boundaries and discriminate categories. Results on standard speech benchmarks show that recurrent, sparse assembly dynamics are promising in low-data settings, while offering a biologically grounded alternative to conventional deep learning approaches.

There are several limitations to our approach. First, the per utterance oracle prominence selection used for boundary detection provides an upper bound (Section~\ref{sec:eval}); using a globally fixed threshold reduces F1 by approximately 0.05--0.10. Second, classification relies on supervised labels to train the per class areas. A fully unsupervised system would require discovering phones categories directly from continuous speech, likely demanding richer and more structured multimodal assembly representations. Developing such mechanisms within AC frameworks for speech remains an important direction for future work. Third, although classification accuracy is well above chance on both tasks (Table~\ref{tab:classification}), it still not as accurate as established deep learning baselines. More expressive input representations or moderately more complex architectures may further improve performance.

Despite these limitations, AC based models hold tremendous promise for the future. Treating speech and audio as trajectories in a dynamical system allows, in principle, unbounded temporal context encoded in the evolving neural state, whereas transformer architectures are constrained by a fixed context window and incur quadratic computational cost with respect to sequence length. To solve a harder problem or take into account a longer context, deep learning systems need to be made bigger. AC based systems allow for the possibility that they just need more time.

\bibliographystyle{IEEEtran}
\bibliography{mybib}

@inproceedings{radford_robust_2022,
	Author = {Alec Radford and Jong Wook Kim and Tao Xu and Greg Brockman and Christine McLeavey and Ilya Sutskever},
	Title = {Robust Speech Recognition via Large-Scale Weak Supervision},
	Booktitle = {Proc. International Conference on Machine Learning (ICML)},
	Year = {2023},
	Pages = {28492--28518}}

@article{baevskiWav2vec20Framework2020,
  Author = {Alexei Baevski and Henry Zhou and Abdelrahman Mohamed and Michael Auli},
  Title = {wav2vec 2.0: A Framework for Self-Supervised Learning of Speech Representations},
  Journal = {Advances in Neural Information Processing Systems (NeurIPS)},
  Volume = {33},
  Pages = {12449--12460},
  Year = {2020}}

@article{hsuHuBERTSelfSupervisedSpeech2021,
  Author = {Wei-Ning Hsu and Benjamin Bolte and Yao-Hung Hubert Tsai and Kushal Lakhotia and Ruslan Salakhutdinov and Abdelrahman Mohamed},
  Title = {{HuBERT}: Self-Supervised Speech Representation Learning by Masked Prediction of Hidden Units},
  Journal = {IEEE/ACM Transactions on Audio, Speech, and Language Processing},
  Volume = {30},
  Pages = {3451--3460},
  Year = {2022},
  doi = {10.1109/TASLP.2022.3193779}}

@article{Davis80-COP,
	Author = {Steven B. Davis and Paul Mermelstein},
	Journal = {IEEE Transactions on Acoustics, Speech and Signal Processing},
	Number = {4},
	Pages = {357--366},
	Title = {Comparison of Parametric Representations for Monosyllabic Word Recognition in Continuously Spoken Sentences},
	Volume = {28},
  Month = aug,
	Year = {1980}}

@article{Rabiner89-ATO,
	Author = {Lawrence R. Rabiner},
	Journal = {Proceedings of the IEEE},
	Number = {2},
	Pages = {257--286},
	Title = {A Tutorial on Hidden {Markov} Models and Selected Applications in Speech Recognition},
	Volume = {77},
  Month = feb,
	Year = {1989}}

@article{TIMIT92,
  Author = {John S. Garofolo and Lori F. Lamel and William M. Fisher and Jonathan G. Fiscus and David S. Pallett and Nancy L. Dahlgren and Victor Zue},
  Title = {{TIMIT} Acoustic-Phonetic Continuous Speech Corpus},
  Journal = {Linguistic Data Consortium},
  Year = {1993}}

@article{Lee89,
  Author = {K.-F. Lee and H.-W. Hon},
  Title = {Speaker-Independent Phone Recognition Using Hidden {Markov} Models},
  Journal = {IEEE Transactions on Acoustics, Speech and Signal Processing},
  Volume = {37},
  Number = {11},
  Pages = {1641--1648},
  Year = {1989}}

@inproceedings{Bergstra11,
  Author = {James Bergstra and R\'{e}mi Bardenet and Yoshua Bengio and Bal\'{a}zs K\'{e}gl},
  Title = {Algorithms for Hyper-Parameter Optimization},
  Booktitle = {Proc. Advances in Neural Information Processing Systems (NeurIPS)},
  Year = {2011},
  Pages = {2546--2554}}

@inproceedings{Akiba19,
  Author = {Takuya Akiba and Shotaro Sano and Toshihiko Yanase and Takeru Ohta and Masanori Koyama},
  Title = {Optuna: A Next-generation Hyperparameter Optimization Framework},
  Booktitle = {Proc. ACM SIGKDD International Conference on Knowledge Discovery and Data Mining},
  Year = {2019},
  Pages = {2623--2631}}

@article{papadimitriouBrainComputationAssemblies2020,
  Author = {Christos H. Papadimitriou and Santosh S. Vempala and Daniel Mitropolsky and Michael Collins and Wolfgang Maass},
  Title = {Brain Computation by Assemblies of Neurons},
  Journal = {Proceedings of the National Academy of Sciences},
  Volume = {117},
  Number = {25},
  Pages = {14464--14472},
  Year = {2020},
  doi = {10.1073/pnas.2001893117}}

@article{dabagiaComputationSequencesAssemblies2024,
  Author = {Max Dabagia and Christos H. Papadimitriou and Santosh S. Vempala},
  Title = {Computation With Sequences of Assemblies in a Model of the Brain},
  Journal = {Neural Computation},
  Volume = {37},
  Number = {1},
  Pages = {193--233},
  Year = {2024},
  doi = {10.1162/neco_a_01720}}

@article{dabagiaAssembliesNeuronsLearn2022,
  Author = {Max Dabagia and Christos H. Papadimitriou and Santosh S. Vempala},
  Title = {Assemblies of Neurons Learn to Classify Well-Separated Distributions},
  Journal = {arXiv preprint arXiv:2110.03171},
  Year = {2022},
  doi = {10.48550/arXiv.2110.03171}}

@article{mitropolskyBiologicallyPlausibleParser2021,
  Author = {Daniel Mitropolsky and Michael J. Collins and Christos H. Papadimitriou},
  Title = {A Biologically Plausible Parser},
  Journal = {arXiv preprint arXiv:2108.02189},
  Year = {2021},
  doi = {10.48550/arXiv.2108.02189}}

@article{weiBionicNaturalLanguage2024,
  Author = {Zhenghao Wei and Kehua Lin and Jianlin Feng},
  Title = {A Bionic Natural Language Parser Equivalent to a Pushdown Automaton},
  Journal = {arXiv preprint arXiv:2404.17343},
  Year = {2024},
  doi = {10.48550/arXiv.2404.17343}}

@article{damorePlanningBiologicalNeurons2022,
  Author = {Francesco D'Amore and Daniel Mitropolsky and Pierluigi Crescenzi and Emanuele Natale and Christos H. Papadimitriou},
  Title = {Planning with Biological Neurons and Synapses},
  Journal = {Proceedings of the AAAI Conference on Artificial Intelligence},
  Volume = {36},
  Number = {1},
  Pages = {21--28},
  Year = {2022},
  doi = {10.1609/aaai.v36i1.19875}}

@article{garagnaniRecruitmentConsolidationCell2009,
  Author = {Max Garagnani and Thomas Wennekers and Friedemann Pulverm{\"u}ller},
  Title = {Recruitment and Consolidation of Cell Assemblies for Words by Way of {Hebbian} Learning and Competition in a Multi-Layer Neural Network},
  Journal = {Cognitive Computation},
  Volume = {1},
  Number = {2},
  Pages = {160--176},
  Year = {2009},
  doi = {10.1007/s12559-009-9011-1}}

@book{hebbOrganizationBehaviorNeuropsychological1949,
  Author = {D. O. Hebb},
  Title = {The Organization of Behavior: A Neuropsychological Theory},
  Publisher = {Wiley},
  Address = {New York},
  Year = {1949}}

@book{quianquirogaPrinciplesNeuralCoding2013,
  Author = {Rodrigo {Quian Quiroga} and Stefano Panzeri},
  Title = {Principles of Neural Coding},
  Publisher = {Taylor \& Francis Group},
  Address = {Boca Raton},
  Year = {2013}}

@article{Warden18,
	Author = {Pete Warden},
	Title = {Speech Commands: A Dataset for Limited-Vocabulary Speech Recognition},
	Journal = {arXiv preprint arXiv:1804.03209},
	Year = {2018}}

@article{Kemker_McClure_Abitino_Hayes_Kanan_2018, title={Measuring Catastrophic Forgetting in Neural Networks}, volume={32}, url={https://ojs.aaai.org/index.php/AAAI/article/view/11651}, DOI={10.1609/aaai.v32i1.11651}, number={1}, journal={Proceedings of the AAAI Conference on Artificial Intelligence}, author={Kemker, Ronald and McClure, Marc and Abitino, Angelina and Hayes, Tyler and Kanan, Christopher}, year={2018}, month={Apr.} }

@ARTICLE{Lake2015-kj,
  title     = "Human-level concept learning through probabilistic program
               induction",
  author    = "Lake, Brenden M and Salakhutdinov, Ruslan and Tenenbaum, Joshua
               B",
  journal   = "Science",
  publisher = "American Association for the Advancement of Science (AAAS)",
  volume    =  350,
  number    =  6266,
  pages     = "1332--1338",
  month     =  dec,
  year      =  2015,
  copyright = "http://www.sciencemag.org/about/science-licenses-journal-article-reuse",
  language  = "en"
}

@ARTICLE{Bi1998-tk,
  title     = "Synaptic modifications in cultured hippocampal neurons:
               dependence on spike timing, synaptic strength, and postsynaptic
               cell type",
  author    = "Bi, G Q and Poo, M M",
  journal   = "J. Neurosci.",
  publisher = "Society for Neuroscience",
  volume    =  18,
  number    =  24,
  pages     = "10464--10472",
  month     =  dec,
  year      =  1998,
  copyright = "https://creativecommons.org/licenses/by-nc-sa/4.0/",
  language  = "en"
}

@ARTICLE{Artola1990-gk,
  title     = "Different voltage-dependent thresholds for inducing long-term
               depression and long-term potentiation in slices of rat visual
               cortex",
  author    = "Artola, A and Br{\"o}cher, S and Singer, W",
  journal   = "Nature",
  publisher = "Springer Science and Business Media LLC",
  volume    =  347,
  number    =  6288,
  pages     = "69--72",
  month     =  sep,
  year      =  1990,
  language  = "en"
}

\end{document}